\documentclass[english,prb,preprint,superscriptaddress]{revtex4}

\usepackage{times}
\usepackage[T1]{fontenc}
\usepackage[latin1]{inputenc}
\usepackage{amsmath}
\usepackage{amssymb}	
\usepackage{MnSymbol}
\usepackage{xfrac}
\usepackage{float}
\usepackage{color}
\usepackage{graphicx}
\usepackage{xfrac}
\makeatletter
\usepackage{babel}
\makeatother

\begin{document}

\title{Modification of electronic surface states by graphene islands on Cu(111)}

\author{S. M. Hollen}\email{hollen.8@osu.edu}
\author{G. A. Gambrel}
\author{S. J. Tjung}
\author{N. M. Santagata}	
\author{E. Johnston-Halperin}
\author{J. A. Gupta}
\affiliation{The Ohio State University}

\begin{abstract} 
We present a study of graphene/substrate interactions on UHV-grown graphene islands with minimal surface contamination using \emph{in situ} low-temperature scanning tunneling microscopy (STM).
We compare the physical and electronic structure of the sample surface with atomic spatial resolution on graphene islands versus regions of bare Cu(111) substrate. We find that the Rydberg-like series of image potential states is shifted toward lower energy over the graphene islands relative to Cu(111), indicating a decrease in the local work function, and the resonances have a much smaller linewidth, indicating reduced coupling to the bulk. In addition, we show the dispersion of the occupied Cu(111) Shockley surface state is influenced by the graphene layer, and both the band edge and effective mass are shifted relative to bare Cu(111). 
 \end{abstract}

\maketitle

Graphene's unique electronic and physical properties have fueled investment for a variety of applications.  However, as it is only one atom thick, graphene's interaction with metal electrodes or substrates can have a strong effect on its inherent properties.\cite{Giovannetti:2008dy}  Copper commands special interest as the most commonly used metal substrate for graphene growth by chemical vapor deposition (CVD).\cite{Li:2009ka}  Copper is also a common contaminant in CVD graphene films transferred to other substrates.\cite{Lin:2012bf}  The electronic states of the (111) face of single crystal copper provide an additional opportunity to probe interfacial electronic structure due to adlayers.  Several recent experiments have taken advantage of these exposed surface states of crystalline metal substrates for such studies.\cite{Feenstra:2013gy,Niesner:2014kp,Craes:2013es,Repp:2004kp}  On Cu(111), an occupied Shockley surface state forms in the projected band gap leading to a quasi-2D electron gas,\cite{Crommie:1993co} which can act as a sensitive probe of changes in the surface electronic potential. Additional unoccupied surface states form due to the potential well created by an electron and its image charge.  These image potential states (IPS) occur in a Rydberg series pinned to the vacuum level and thus serve as a local probe of the work function. Studies of Shockley states and IPS can help determine the interaction between adlayers and substrates as well as the resulting charge transfer to the adlayer.\cite{Giovannetti:2008dy}  Most of these studies have involved either thin dielectric films\cite{Repp:2004kp}, noble gases\cite{Park:2000cm}, few-layer metals,\cite{Malterre:2007hc,diekhoner_surface_2003} or self-assembled molecular monolayers.\cite{Heinrich:2011du}   As graphene is a metallic 2D crystal with weak van der Waals substrate coupling, surface state changes due to a graphene sheet may lead to new understanding of graphene/metal interactions as well as surface state physics. The graphene/metal system also presents an interesting opportunity to study whether a 2D metal that is entirely surface will present its own surface states, as was recently predicted for freestanding graphene.\cite{Silkin:2009fr} 
 
Recently, the surface states and electronic structure of graphene on a number of metal substrates have been studied using photoemission spectroscopy.\cite{Nobis:2013he, Walter:2011fj,Jeon:2013fw, Kong:2010hr} Scanning tunneling microscopy (STM) and spectroscopy (STS) studies provide a complementary viewpoint as a local probe sensitive to both occupied and unoccupied electronic states. In these experiments local topographical images provide structural information of measured sites, and local spectroscopy avoids the complexity of aggregated rotational domains.  Detailed STM studies of IPS have been carried out for graphene on Ru(0001),\cite{Feng:2011vb} SiC,\cite{Bose:2010fd} and Ir(111),\cite{Craes:2013es} but have not been reported for Cu(111). For graphene on Cu(111),  previous STS work has primarily been focused on characterizing the UHV growth,\cite{Gao:2010tl,Zhao:2011gc,Jeon:2011dp} and a detailed study of the interactions between the Cu Shockley state and graphene is lacking.  Here we use a reproducible ultra high vacuum (UHV) CVD process for growing pristine graphene islands on Cu(111) to study the electronic structure of the resulting gr/Cu heterostructure.  Our STM and STS measurements indicate that the graphene islands significantly alter the IPS and Shockley surface states of the Cu.  The graphene layer reduces the work function, decouples the IPS from the Cu bulk states, and reduces the effective mass of the electrons in the Shockley surface state. 

Graphene islands are grown on a clean Cu(111) single crystal by heating the crystal near 1000$^\circ$C in the presence of $10^{-5}$ mbar ethylene gas in four 5 minute temperature cycles (for details, see Supp. Mat\cite{SuppMat}). This procedure results in an approximately 1/5 monolayer coverage of graphene islands surrounded by clean Cu(111). The islands either grow continuously over, or terminate at Cu step edges (Fig. 1a,b,d), and often have hexagonal shapes (Fig. 1e,f). At low tip-sample biases, scattering of the Shockley surface state electrons is observed as standing waves\cite{Crommie:1993co} in images of bare Cu and gr/Cu regions (Fig. 1a,f). The point defect density is much lower in graphene regions; defects concentrate at the gr/Cu boundary as if the growing graphene island swept them clean (see \emph{e.g.} Fig. 1d). Defects in the graphene lattice appear as bright spots (\emph{e.g.} Fig. 1e) or dark lines (Fig. 1c). An atomically-resolved image of the bright defects on a step edge reveals a triangular structure (upper terrace of Fig. 1a and Fig. 1b) consistent with a single carbon-site graphene defect.\cite{Zhao:2011ft, Amara:2007vh, Wehling:2007bx}  Atomic resolution images of the dark lines (not shown) identifies them as graphene grain boundaries, which sometimes have a hexagonal shape (\emph{cf.}, yellow arrow in Fig. 1c). The continuity of the graphene sheet over step edges is demonstrated by the continuous moir\'e pattern between the graphene and Cu crystal (Fig. 1d), and atomic resolution images of the graphene lattice on step edges (Figs. 1a,b). Islands exhibit a variety of moir\'e patterns with lattice spacings and corresponding orientation angles (relative to Cu) including 1.6 nm/9$^\circ$, 2.0 nm/7$^\circ$, 3.0 nm/4$^\circ$, 4.5 nm/2$^\circ$, and 6.5 nm/0$^\circ$. The continuous growth of graphene over step edges and the observation of a number of moir\'e patterns suggest that the graphene does not have a strong chemical interaction driving alignment with the underlying Cu.  

The pristine, contamination-free surface of the graphene islands provides an excellent opportunity to compare the surface electronic structure of clean Cu and gr/Cu surfaces with tunneling spectroscopy (\emph{cf.}, Methods).  We schematically summarize our findings on the electronic structure of graphene-covered Cu(111) and compare to bare Cu(111) in Fig. 2a. At the Cu(111) surface, the Rydberg series of IPS form near the vacuum level beginning with the $n=1$ at $\sim$5 eV, and the Shockley surface state forms in the projected bulk band gap, $\sim$0.4 eV below the Fermi energy (right side of Fig. 2a). We find that the graphene layer reduces the work function, shifts both the IPS and the Shockley surface state toward the Fermi energy, and reduces the effective mass of electrons in the Shockley state (left side of Fig. 2a). In the following, we discuss each of these measurements in detail. 

We probe the difference in work function, $\Delta\Phi$, between Cu and gr/Cu surfaces by measuring the apparent barrier height, $\Phi_{a}$, using $\frac{d\ln{I}}{d\Delta{z}}=\frac{2\sqrt{2m\Phi_{a}}}{\hbar}$, where  $m$ is the electron mass.\cite{SuppMat}  Fig. 2b shows measurements taken with the same microscopic tip, giving $\Phi_a =$ 4.4 eV and 3.5 eV on Cu and gr/Cu, respectively.  While the absolute work function depends on the barrier approximation and detailed information about the tip shape and work function, the difference in apparent barrier height represents a quantitative difference in surface work function.  Using a simple trapezoidal approximation to the tunneling barrier: $\Phi_{a}\approx\frac{\Phi_t+\Phi_s}{2}$, where $\Phi_t$ and $\Phi_s$ are the tip and sample work functions, respectively,\cite{SuppMat} we find $\Delta\Phi \approx$ 1.8 eV.  This work function difference results from the dipole that forms at the gr/Cu interface and indicates an electric field directed towards the copper. It represents a relatively large change in $\Phi$ compared to measurements of alkalis on metals,\cite{Zangwill} though we note that the barrier approximation would tend to overestimate the difference in work function. Comparisons with the literature can be complicated by experimental variation, including sample exposure to air, which causes shifts in work function due to intercalants, and approximations due to experimental technique, such as the barrier estimation required here.  Overall, the change we report qualitatively agrees with theoretical expectations\cite{Giovannetti:2008dy} and low energy electron reflectivity results,\cite{Srivastava:PRB2013} which report a 0.9 eV and 0.82 eV decrease in work function on gr/Cu, respectively. \emph{A priori}, the reduction in work function due to the graphene island seems at odds with the n-doping indicated by photoemission,\cite{Walter:2011fj,Kong:2010hr} but may highlight the distinction between the overall charge transfer and the interfacial dipole responsible for the work function change.

We now present tunneling spectroscopy measurements of the island in Fig. 2c, which are characteristic of the gr/Cu surface. STS measurements probing the IPS and Shockley surface states are taken separately because of their different energy ranges, as described in Methods. The IPS produce tunneling resonances at bias voltages within the field emission regime. These resonances are Stark shifted by the electric field in the tunnel junction,\cite{Binnig:1985bz} and thus do not directly correspond to the intrinsic IPS measured with photoemission and schematically depicted in Fig. 2a. Fig. 2d compares these resonances taken with the tip positioned over bare Cu and gr/Cu areas, as marked in Fig. 2c.  In the constant current mode of STS (\emph{cf.} Methods), the IPS resonances produce sharp peaks in $(dI/dV)\vert_I$ and corresponding steps in tip height $z(V)$.  IPS in gr/Cu regions appear at lower energies than their Cu counterparts. Within the voltage range studied, four IPS are observed on Cu, with the $n=1$ state at 4.8 V, and five IPS are observed on gr/Cu, with the $n=1$ state at 4.2 V. The shift in $n=1$ IPS is consistent with the reduction in work function indicated by the apparent barrier height measurements in Fig. 2b.  This shift is reproducible over many different regions of the sample and with many different microscopic tips (see statistics in Supp. Fig. 2\cite{SuppMat}). 

Also notable are the widths of the IPS resonances, which reflect the lifetime of an electron in the image potential state.\cite{Dougherty:2007cu,Crampin:2005fy}  We find the IPS lifetimes on bare Cu, $\tau_{Cu}\sim\hbar/\Delta E= 0.8$ fs for the $n=1$ IPS in Fig. 2d, are significantly shorter than those on graphene, $\tau_{gr/Cu}\sim 2$ fs.   This increase in lifetime indicates that the IPS over gr/Cu regions are significantly more decoupled from bulk Cu states, consistent with photoemission measurements of graphene on other metal surfaces.\cite{Nobis:2013he}  This decoupling can be qualitatively understood by considering the effect of graphene on the surface potential and the available bulk states.  The decreased surface potential physically shifts the IPS further from the surface, and reduces wavefunction overlap with bulk states. It also shifts the energy of the IPS further into the surface-projected band gap, reducing degeneracy with unoccupied bulk Cu states.  Finally, we note that we observe spatially-dependent splitting in the $n=1$ IPS state over graphene islands.  As discussed in Supp. Mat.,\cite{SuppMat} this splitting can either be attributed to quantum confinement effects by narrow graphene regions or to the hybridization of IPS states inherent to the graphene sheet\cite{Silkin:2009fr} with those associated with the Cu. 

Turning to the Shockley surface state, STS measurements reveal a change in the surface electronic potential over the graphene islands. STM images at low bias reveal standing wave patterns of electron scattering in both the bare Cu and gr/Cu regions. Figure 2c illustrates these standing waves, originating from scattering at step edges and gr/Cu boundaries. Some point defects also scatter the surface state electrons, such as the two dark defects at the right of the gr/Cu region of Fig. 2c. In contrast, triangular graphene lattice defects do not scatter the surface state electrons (there are two in Fig. 2c that are not resolved at this bias voltage). These differences reflect the degree to which the electrostatic environment is perturbed by the defect, as recently observed during the switching of adatoms on NaCl/Cu(111) between neutral and charged states.\cite{Repp:2004kp} The Shockley state band edge is evident in tunneling spectroscopy as a step in the dI/dV signal. Figure 4a shows that the Shockley band edge appears at -0.44 V with the tip positioned over bare Cu, and shifts to  -0.32 V in the gr/Cu regions. Since the position of the band edge is determined by the electrostatic potential at the surface, the formation of an interface dipole between the Cu substrate and the graphene (as indicated by the reduction in work function) causes this shift.  The observed direction of the shift is consistent with expectations for a potential created by an interface dipole with positive charge near the graphene.\cite{Khomyakov:2009fs}   

We next explore the energy dependence of the scattering of Shockley electrons to measure their dispersion in gr/Cu versus Cu regions. We extract the characteristic scattering wavelength of these electrons at discrete energies from spatial maps of the $dI/dV$ signal (Fig. 3b-e) and plot dispersion in Fig. 3f. Performing a fast Fourier transform (FFT) of the $dI/dV$ maps results in a circular ring with a radius proportional to the crystal momentum.\cite{Petersen:2000hb}  Because the maps show the local electron density, the wavelength indicated by the FFT is actually half the wavelength of the electrons. The crystal momentum for each energy is determined from a Lorentzian fit to the radial average of these FFT. The data are analyzed using a parabolic 2D free-electron-like model $E(k_\parallel)=\frac{\hbar^2}{2m^*}k_\parallel^2+E_0$, where $m^*$ is the effective mass and $E_0$ is the Shockley band edge. The resulting dispersions for the surface state in gr/Cu and bare Cu are compared in Fig. 3f. On bare Cu, we note that for a range of bias voltages 0-200 mV ($E-E_0=$422-622 meV) scattering into bulk states is observed as an additional scattering ring.\cite{Schouteden:2009hh,Reinert:2001er} This bulk channel is greatly suppressed on gr/Cu, where only one ring is observed.   In both cases, the data are fit well by the free electron model, which yields effective electron masses for Cu and gr/Cu of $(0.420\pm0.009\pm0.008)m_e$ and $(0.386\pm0.009\pm0.007)m_e$, respectively. Here the first uncertainty is derived from the parabolic fit while the second estimates systematic errors such as drift during imaging. 

Previous studies of graphene on Cu(111) and Au(111) substrates also found an upward shift in the Shockley band edge.\cite{Walter:2011fj,Batra:2014gca}  However, $m^*$ was either found to be unchanged within the experimental resolution\cite{Jeon:2011dp,Walter:2011fj}, or in the case of graphene nanoribbons on Au(111), increased.\cite{Batra:2014gca}  In fact, the combination of a shift in the band edge toward $E_F$ and a decrease in $m^*$ observed for the gr/Cu(111) system here runs counter to the typical trend observed.  For dielectric overlayers,\cite{Repp:2004kp} noble gases,\cite{Andreev:2004eg} self-assembled molecular monolayers,\cite{Heinrich:2011du} Ag on Au(111),\cite{Malterre:2007hc} and monolayer BN\cite{Joshi:2012ej} an upward shift in the Shockley band edge is accompanied by an increased $m^*$. The physics underlying the trend must relate the interface dipole, which shifts the band edge, to the electronic surface corrugation, which modifies the effective mass of surface state electrons.\cite{SMITH:1985uu}  Here we find that the surface dipole shifts the band edge up in energy while an increased surface corrugation reduces $m^*$.  This observation may be similar to Co overlayers on Cu(111),\cite{diekhoner_surface_2003} where $m^*$ was also found to be decreased by Co while the Shockley band edge increased. 

In conclusion, we measured the modification of electronic surface states by nanometer sized graphene islands on a Cu(111) crystal with atomic spatial resolution.  Using multiple spectroscopic techniques, we found a work function reduction of 1.8 eV over the graphene islands, and a concomitant shift in the image potential states.  This shift and an increase in the IPS electron lifetime shows that graphene acts to decouple these surface states from the bulk Cu states. Finally, we find that the effective mass of Shockley surface state electrons is reduced in the gr/Cu regions.  Together, these results show that the interaction between the graphene and the Cu significantly alters the electronic behavior at the surface.  The electrostatic surface potential created by the gr/Cu stack influences the electronic states in the graphene as well as the charge transfer between the materials.  These interactions can dramatically affect the mobility and transport across interfaces, which has strong implications for graphene devices.

 \begin{acknowledgements} 
Funding for this research was provided by the Center for Emergent Materials at the Ohio State University, an NSF MRSEC (Award Numbers DMR-1420451 and DMR-0820414).  We would also like to thank Leonard Brillson for helpful discussions.
\end{acknowledgements}


\clearpage
\begin{figure}
\begin{center}
\includegraphics[width=0.8\columnwidth,keepaspectratio]{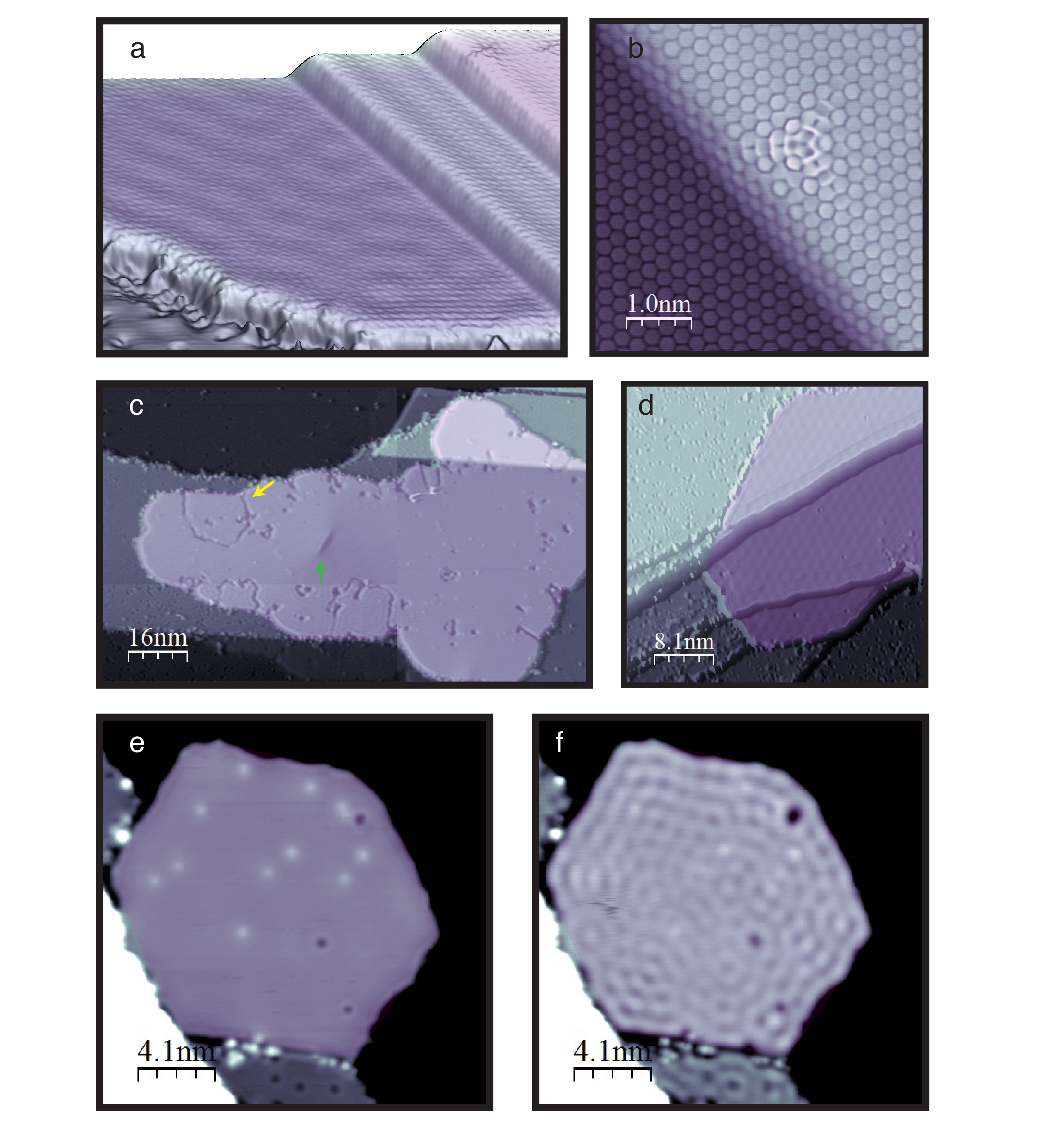}
\caption{Graphene islands (colored lavender) grown on Cu(111) (gray) a) 3D view of an atomically resolved graphene sheet draped over 2 Cu step edges.  Two graphene lattice defects are visible on the top Cu terrace (upper right).  The graphene/Cu boundary is shown in the lower left. (Image taken at V=-10 mV, I=0.5 nA) (20x20 nm) b) Atomically resolved graphene lattice defect in a graphene sheet draped over a Cu step edge. (10 mV, 0.2 nA) c) Mosaic image of a large island with grain boundaries appearing as dark lines.  Yellow arrow: hexagonal grain boundary.  Green arrow: unidentified defect. (1 V, 0.2 nA)   d) Continuous growth of a graphene island over 3 Cu terraces.  Moir\'e pattern with 1.6 nm periodicity indicates an angle of 9$^\circ$ between the graphene and Cu lattices. (1 V, 0.5 nA)  e,f) Hexagonal graphene island at the edge of a Cu terrace with 3 dark defects (likely CO) and 12 graphene lattice defects (bright spots in (e)). Imaged at 1 V and 1 nA (e) and 50 mV and 0.2 nA (f).  Image (b) is a topographic image superimposed with its Laplacian; images (c) and (d) are topographic images superimposed with their derivatives. These retain topographic information and accentuate small changes in height.
\label{cap:fig1}}
\end{center}
\end{figure}

\begin{figure}
\begin{center}
\includegraphics[width=1\columnwidth,keepaspectratio]{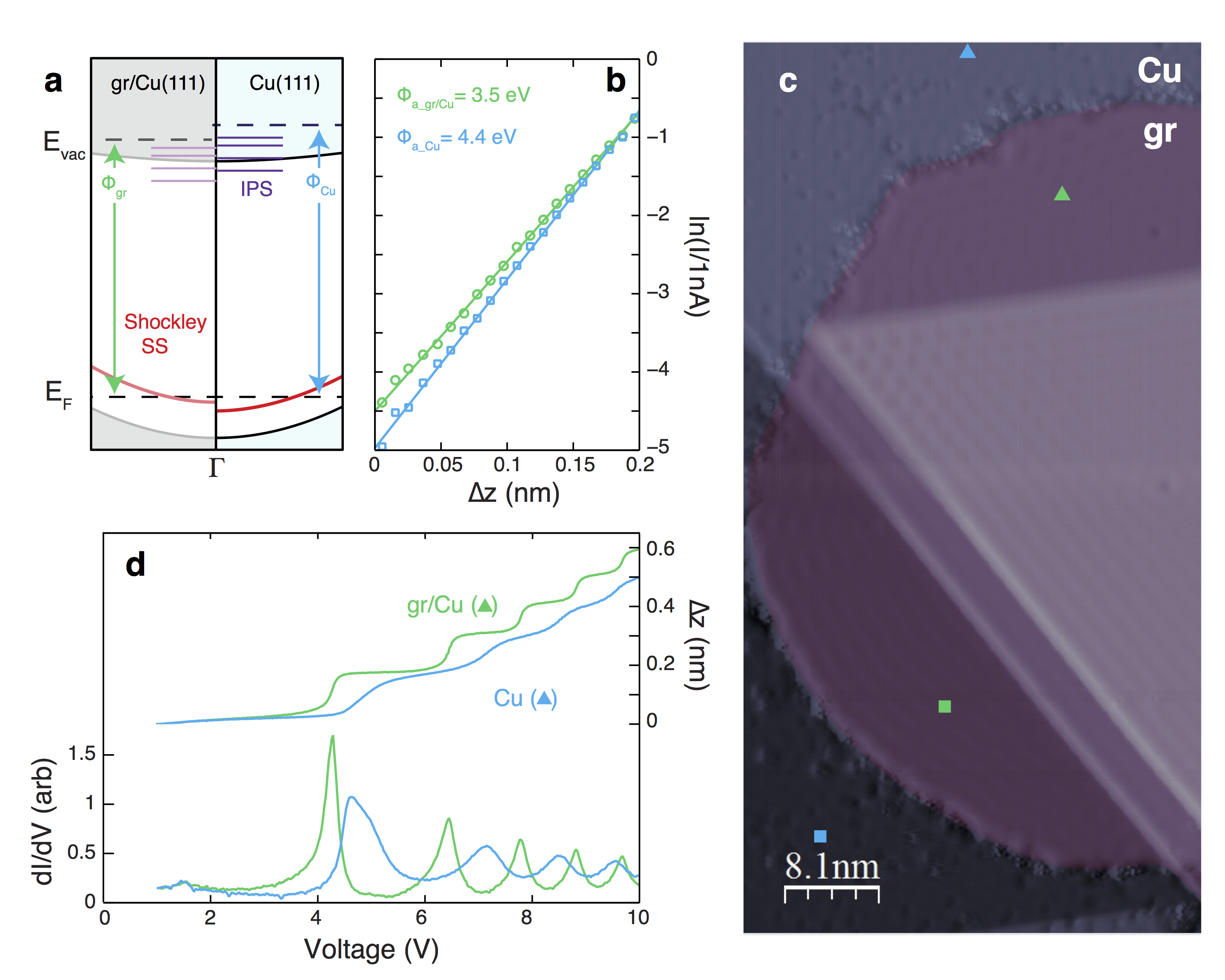}
\caption{Electronic surface states: Cu(111) vs. graphene/Cu(111) a) Schematic energy-momentum diagram of surface states in Cu(111) (right half) and the observed changes for graphene covered Cu(111) (left half). The shaded area represents the projected bulk states for Cu(111).  The Fermi and the vacuum energies ($E_{F}$ and $E_{vac}$, respectively) are marked with horizontal dashed lines.  The Shockley surface state is shown as a red parabola. A series of image potential states is shown in blue near $E_{vac}$.  For gr/Cu, only the band bottom is marked (horizontal dashed lines) as we did not measure their dispersion.  $\Phi_{Cu}$ and $\Phi_{gr}$ depict the work functions of Cu(111) and gr/Cu, respectively. b) Tunnel current on a log scale ($\ln(I/1 \mathrm{nA})$) versus change in tip height ($\Delta Z$) at constant bias voltage (0.1 V) with the tunneling conditions set at 0.5 nA and 0.1V. Apparent tunnel barrier height of 4.4 eV for Cu(111) (blue) and 3.5 eV for gr/Cu (green) is extracted from the slopes of the linear fit to the curves according to Eqn. 1.  c) STM image of graphene island draped over Cu(111) step edges (2 image mosaic). (50 mV, 1 nA) d) $(dI/dV)\vert_I$ showing resonant tunneling into image potential states on Cu (blue) and gr/Cu (green) (at 1 nA). Triangles in (c) indicate the tip position.
\label{cap:fig2}}
\end{center}
\end{figure}

\begin{figure}
\begin{center}
\includegraphics[width=1\columnwidth,keepaspectratio]{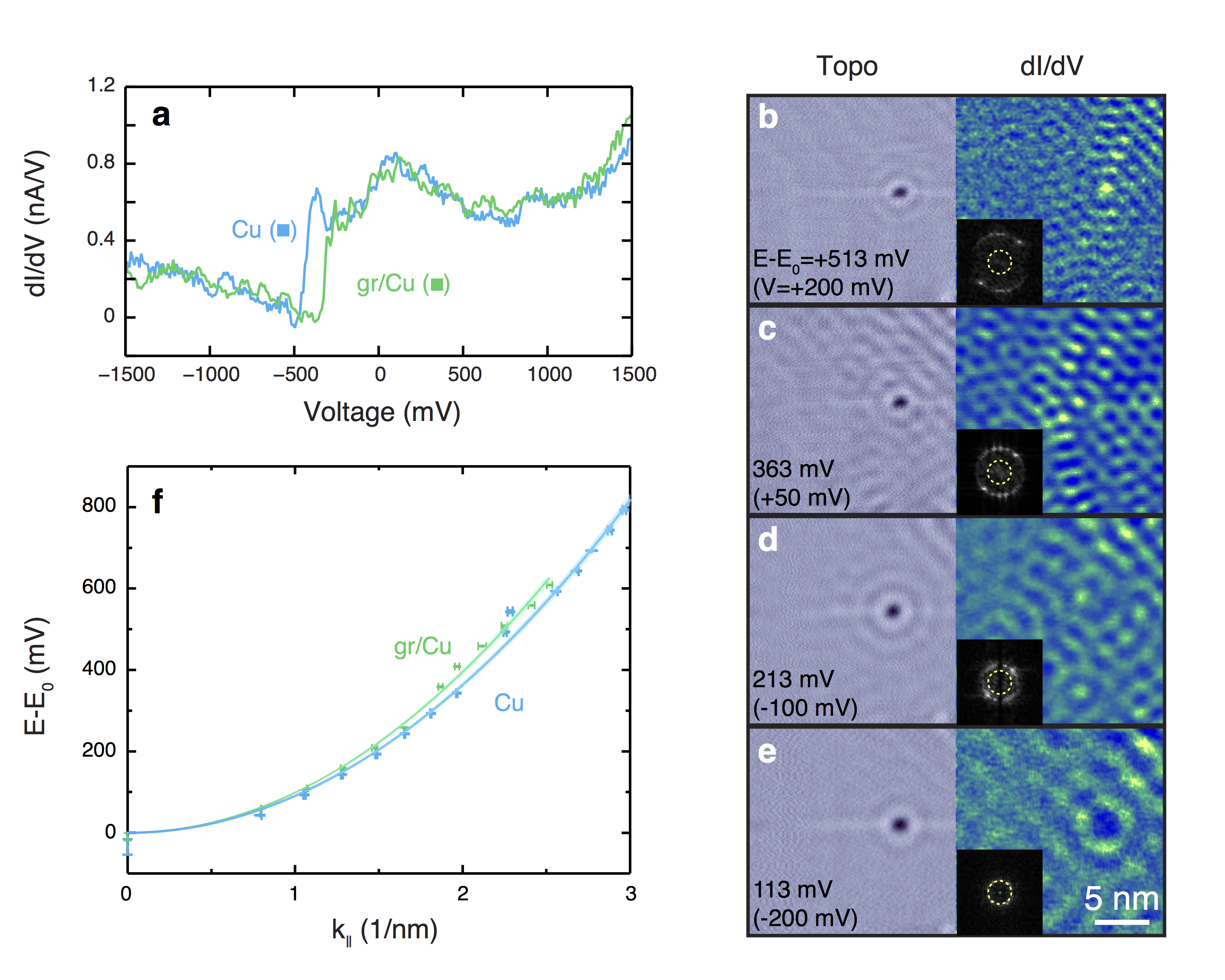}
\caption{Graphene modification of Cu(111) Shockley State
a) Point spectra on clean Cu(111) (blue) and gr/Cu (green) for points in Fig. 2c (squares).  Shockley band edge appears as a sharp step at -0.44 V (Cu(111)) and -0.32 V (gr/Cu).
b-e) Representative topographic images (left) and simultaneously acquired differential conductance maps (right) of a gr/Cu region at various bias voltages ($I = 0.5 nA$). The insets show FFTs of the dI/dV maps (see text). The yellow dashed circle has a radius of $k=$ 1 1/nm.  
f) Dispersion of the Shockley surface state for graphene covered Cu(111) (green) and clean Cu(111) (blue) extracted from radial averages of FFTs. The data are plotted from the respective band edges for Cu (-0.46 V) and gr/Cu (-0.32 V) to emphasize the difference in curvature indicating a lower effective mass for gr/Cu states. Error bars report random error and shaded curves represent systematic uncertainty arising from piezo calibration and thermal drift. 
\label{cap:fig4}}
\end{center}
\end{figure}

\end{document}